\documentclass{aa}
\usepackage{graphicx}
\usepackage[]{hyperref}
\hypersetup{colorlinks=true,citecolor=blue}
\usepackage{lipsum, babel}
\usepackage{booktabs}
\usepackage{comment}
\usepackage{txfonts}
\usepackage{enumitem} 

\usepackage{caption}
\usepackage{subcaption}
\usepackage[dvipsnames]{xcolor}
\usepackage{colortbl}

\usepackage{adjustbox}

\usepackage{tablefootnote}
\usepackage{threeparttable}

\def\coo{CO$_2$\xspace}

\def\hh{H$_2$\xspace}

\def\hho{H$_2$O\xspace}
\def\chhhh{CH$_4$\xspace}

\def\hhco{H$_2$CO\xspace}

\def\chhhoh{CH$_3$OH\xspace}

\def\cchhhh{C$_2$H$_4$\xspace}

\def\chhhcho{CH$_3$CHO\xspace}

\def\nhhh{NH$_3$\xspace}

\def\Rearth{$R_\oplus$\xspace}

\begin{document}

   \title{Unraveling the non-equilibrium chemistry of the temperate sub-Neptune K2-18\,b}
   
   \titlerunning{Non-equilibrium chemistry of K2-18\,b}

   \author{A. Y. Jaziri\inst{1,2}, O. Sohier\inst{1}, O. Venot\inst{3} \and N. Carrasco\inst{1,4}}
    
   \authorrunning{Jaziri et al.}

   \institute{$^1$ LATMOS/IPSL, UVSQ Universit\'{e} Paris-Saclay, Sorbonne Universit\'{e}, CNRS, Guyancourt, France \\
   $^2$ Laboratoire d’astrophysique de Bordeaux, Univ. Bordeaux, CNRS, B18N, all\'{e}e Geoffroy Saint-Hilaire, 33615 Pessac, France \\
   $^3$ Universit\'{e} Paris Cit\'{e} and Univ Paris Est Creteil, CNRS, LISA, F-75013 Paris, France \\
   $^4$ ENS Paris-Saclay, Universit\'{e} Paris-Saclay, Gif-sur-Yvette, 91190, France \\
}

   \date{Received 12 May 2025; accepted 17 July 2025}

 
  \abstract
   {The search for habitable, Earth-like exoplanets faces significant observational challenges due to their small size and faint signals. M-dwarf stars provide an opportunity to detect and characterize smaller planets, particularly sub-Neptunes, which are among the most common exoplanetary types. K2-18\,b, a temperate sub-Neptune in the habitable zone of an M-dwarf star, has been studied using HST and JWST, revealing an \hh-rich atmosphere with detected \chhhh and possible \coo. However, previous debates in atmospheric composition, emphasize the importance of non-equilibrium chemistry models. These models are crucial for interpreting exoplanetary atmospheres, constraining key parameters such as metallicity, C/O ratio, and vertical mixing (K$_{zz}$).}
   {This study aims to comprehensively explore the parameter space of metallicity, C/O ratio, and K$_{zz}$ for K2-18\,b using the non-equilibrium chemical model \textit{FRECKLL} in conjunction with JWST observational data. By refining these constraints, we seek to improve our understanding of the planet’s atmospheric composition and detect minor species in a temperate sub-Neptune within the habitable zone of an M-dwarf star.}
   {Our approach involves running non-equilibrium chemical models across a three-dimensional parameter space (metallicity, C/O ratio, and K$_{zz}$), generating the corresponding theoretical spectra, and comparing these spectra to JWST observational data to refine atmospheric constraints. This approach assumes a fixed P–T profile, which is sufficient to capture first-order chemical trends, though it introduces some uncertainty in the derived values.}
   {We retrieved a best-fit atmospheric model for K2-18\,b favoring high metallicity (266$^{+291}_{-104}$ at 2$\sigma$) and a high C/O ratio (C/O $\ge$ 2.1 at 2$\sigma$). \chhhh is robustly detected (log$_{10}$[CH$_4$] = -0.3$^{+0.1}_{-1.7}$ at 1 mbar), while \coo and other species remain uncertain due to observational noise and spectral overlap. The K$_{zz}$ has no significant impact on the fit and remains unconstrained. Non-equilibrium models exceed 4$\sigma$ confidence over a flat-line which validates the presence of atmospheric features. Several minor species may exist at ppm levels, though their features are likely masked by dominant species.}
    {We used non-equilibrium chemical models and JWST data to investigate the atmosphere of K2-18\,b, revealing a high metallicity, a high C/O ratio and complex chemical composition. While \chhhh is robustly detected, CO$_2$ remains uncertain, and minor species like H$_2$O and NH$_3$ are likely present. A lower limit on the C/O ratio is constrained, though no upper limit is established. The high C/O ratio also suggests a higher probability of aerosol formation. Our findings highlight the limitations of traditional retrievals with constant abundances and the importance of non-equilibrium models with combining exploration on a large range of metallicity and C/O ratio. Although equilibrium models can reproduce the data, physical conditions indicate that the atmosphere is in a non-equilibrium state, highlighting the limited constraints from current observations and the pressing need for improved data. Future observations with JWST NIRSpec G395H and ELT/ANDES will be key to refining atmospheric constraints and probing potential habitability.}


   \maketitle
%

\section{Introduction}

The search for Earth-like habitable planets is challenging due to their small size and low brightness, which reduce the signal-to-noise ratio and strain current detection and characterization techniques. A few candidates have been identified by leveraging the smaller size and lower luminosity of M-dwarf stars, such as those in the TRAPPIST-1 system \citep{Gillon2017}. However, none have been confirmed to host an atmosphere detectable with current instruments.

Despite this, M-dwarf stars, which constitute the majority of stars in the galaxy, provide a unique opportunity to detect and characterize smaller exoplanets. In particular, sub-Neptunes (1.7 \Rearth < R < 3.5 \Rearth) orbiting M-dwarf stars are ideal candidates for study. Their H$_2$-rich atmospheres result in large scale heights, leading to prominent spectral features that are more easily detectable. This remains true even for temperate sub-Neptunes such as K2-18\,b \citep{Benneke2017} and TOI-270 d \citep{Gunther2019}. Sub-Neptunes are among the most commonly detected exoplanets and thus dominate the known population \citep{Fressin2013, Luger2015, Owen2016, VanEylen2018, Hardegree2018}. They are particularly intriguing because they have no direct analogs in our solar system, raising fundamental questions about the processes that shape their atmospheres, interiors, formation, and potential habitability \citep{Cabot2024}. The habitability of such planets orbiting M-dwarf stars is actively being investigated and is challenged by unique environmental conditions such as intense stellar flares, high-energy particle events, and limited ultraviolet flux \citep{Scalo2007,Ricker2015,Shields2016,Rimmer2018}.

K2-18\,b is a temperate sub-Neptune orbiting within the habitable zone of an M-dwarf star. Its atmosphere was first characterized using the Hubble Space Telescope (HST) \citep{Tsiaras2019, Benneke2019} and more recently with the James Webb Space Telescope (JWST) \citep{Madhusudhan2023, Wogan2024, Schmidt2025}. HST’s Wide Field Camera 3 (WFC3) provided a limited wavelength range (1.1–1.7 $\mu$m) to analyze the planet’s atmospheric composition. While this was sufficient to confirm an H$_2$-dominated atmosphere, it was not enough to distinguish between the presence of H$_2$O \citep{Tsiaras2019, Benneke2019} and CH$_4$ \citep{Blain2021}. JWST, with its broader wavelength coverage (0.9–5.2 $\mu$m) enabled by the combined NIRISS SOSS and NIRSpec G395H instruments, resolved this ambiguity in favor of CH$_4$, detecting approximately 1\% CH$_4$ and 1\% CO$_2$ \citep{Madhusudhan2023}. \cite{Blain2021} had previously predicted the presence of CH$_4$ using disequilibrium chemistry models, emphasizing the crucial role of such models in accurately interpreting exoplanetary atmospheres. Under these temperature, irradiation, and transport conditions, non-equilibrium chemistry dominates over equilibrium chemistry. This necessity is further reinforced by the recent detection of SO$_2$, a photochemical product, in the atmosphere of WASP-39 b \citep{Rustamkulov2023, Alderson2023, Tsai2023}. Non-equilibrium chemistry models applied to K2-18\,b have also led to a reassessment of the retrieved abundances of CH$_4$ and CO$_2$, favoring a higher CH$_4$ content and a lower CO$_2$ content \citep{Wogan2024}. Additionally, a reanalysis of JWST data reduction has cast doubt on the initial CO$_2$ detection itself \citep{Schmidt2025}, aligning with the findings from non-equilibrium chemistry models. This highlights the importance of accounting for non-equilibrium processes in atmospheric models and applying diverse data analysis techniques to ensure robust results, as emphasized by \cite{Schmidt2025}.

Furthermore, the relatively low resolution of space-based observations compared to ground-based ones limits detection to only the major atmospheric constituents, making it difficult to detect minor gases. Chemical models are essential for inferring full atmospheric composition. However, such models still require observational constraints, particularly on metallicity and the C/O ratio. Previous studies using disequilibrium models have mainly explored metallicity, spanning a range from 1 to 1000 times the solar value \citep{Charnay2021, Blain2021, Bezard2022}. Analyses based on HST data \citep{Tsiaras2019, Benneke2019} estimated metallicity between 100 and 300 times solar, consistent with \cite{Fortney2013}, who predicted metallicity of approximately 100 to 400+ times solar for planets with radii of $\sim$2–4 Earth radii. Additionally, \cite{Bezard2022} explored eddy diffusion coefficients (K$_{zz}$) ranging from 10$^5$ to 10$^{10}$ cm$^2$.s$^{-1}$, finding the best fit at 10$^8$ cm$^2$.s$^{-1}$. The C/O ratio is often assumed to be solar, as in \cite{Wogan2024}, though \cite{Blain2021} proposed exploring a small range from 0.0013 to 2.2, with results favoring moderate C/O ratios. More recently, \citet{Schmidt2025} reanalyzed JWST data and investigated disequilibrium models with a metallicity of 100 times solar, K$_{zz}$ of 10$^7$ cm$^2$.s$^{-1}$, and a C/O ratio ranging from solar to 4. They suggested that a high C/O ratio ($\gtrsim$3) may explain the non-detection of H$_2$O.

In this study, we aim to comprehensively explore the three key parameters simultaneously (metallicity, C/O ratio, and K$_{zz}$) using the non-equilibrium chemical model \textit{FRECKLL} \citep{alrefaie2024} together with JWST observational data from \citet{Madhusudhan2023}. Our goal is to better constrain these parameters and investigate the chemistry of minor species in K2-18\,b, a temperate sub-Neptune located in the habitable zone of its host star.


\section{Methods}

Our method consists of three steps: running non-equilibrium chemical models across a three-dimensional parameter space (metallicity, C/O ratio, and K$_{zz}$), computing the corresponding theoretical spectra, and comparing these spectra to the observational data.

\subsection{Observational data}

We used the K2-18\,b observational data from \cite{Madhusudhan2023}. The spectrum was obtained from a single transit observation with each of JWST’s NIRISS SOSS and NIRSpec G395H instruments, covering a discontinuous wavelength range from 0.9 to 5.2 $\mu$m. As retrieved by \cite{Madhusudhan2023}, the spectra exhibit an offset of -41 ppm between the NIRISS and NIRSpec data. For model comparison, we use the spectra at their native resolution and apply the $\chi^2$ method. However, for visualization purposes, we present the lower-resolution spectra from \cite{Madhusudhan2023}.

\subsection{Equilibrium chemistry retrieval}

Even though temperature, irradiation, and transport conditions indicate that non-equilibrium chemistry dominates over equilibrium chemistry, we performed a Bayesian retrieval assuming chemical equilibrium. The objective was to assess whether the current data can distinguish between equilibrium and non-equilibrium chemical states. For this purpose, we used \textit{TauREx 3} (Tau Retrieval for Exoplanets)\footnote{\url{https://github.com/ucl-exoplanets/taurex3}}, a fully Bayesian inverse atmospheric retrieval framework \citep{Al-Refaie2021}.

\textit{TauREx 3} comprises two main components: the Forward Model framework and the Retrieval framework. The Retrieval framework aims to fit a forward model to observational data. As our equilibrium chemical model, we employed \textit{ACE} (coupled with \textit{TauREx 3}) \citep{Agundez2012, Agundez2020}, which has been identified by \cite{Jaziri2024} as the most suitable equilibrium chemistry model for our case.

We used absorption cross-sections computed by ExoMol \citep{Yurchenko2011, Tennyson2012, Barton2013, Yurchenko2014, Barton2014}, specifically those from \cite{Chubb2021}, for the following species: H$_2$O, CO, CH$_4$, CO$_2$, H$_2$CO, HCN, C$_2$H$_2$, NH$_3$, and C$_2$H$_4$. The temperature profile was fixed to match that used in the non-equilibrium chemical models. All free parameters and priors are shown in Table \ref{tab: priors}

\begin{table}[!htbp]
    \caption{Free parameters and priors for the retrieval with equilibrium chemistry.}
    \centering
    \begin{tabular}{@{}lc@{}}
    \toprule
    \toprule
    \textbf{Parameters}           & \textbf{Bounds}  \\ \midrule
    log$_{10}$(P$_{clouds}$) [Pa] & -2 to 6          \\
    radius [R$_{jup}$]            & 0.1 to 0.3       \\ \midrule
    \multicolumn{2}{c}{Chemistry}                    \\ \midrule
    log$_{10}$[Metallicity]       & -1 to 3          \\
    C/O                           & 0.01 to 10       \\ \bottomrule
    \end{tabular}
    \label{tab: priors}
\end{table}

After obtaining the best-fit model with \textit{TauREx 3}, we calculated its $\chi^2$ value, using the same procedure as for the non-equilibrium models, in order to directly compare the two approaches and evaluate whether the available data significantly favor non-equilibrium chemistry over equilibrium chemistry.

\subsection{Non-equilibrium chemical model}

To fit the observed spectrum and determine the chemical composition of the atmosphere of K2-18\,b, we use the 1D non-equilibrium chemical model \textit{FRECKLL} \citep{alrefaie2024}. \textit{FRECKLL} solves the steady-state continuity equations for each species, accounting for chemical production and loss as well as vertical transport via eddy diffusion. The chemical network includes $\sim$2000 reactions (thermal and photochemical) among over 100 neutral species, involving the elements H, C, N and O. The reaction network is primarily adapted from the \cite{Venot2012} and \cite{Moses2011} frameworks, with updates to reaction rates and species relevant to temperate sub-Neptunes. We used specifically the \cite{Venot2020a} chemical network, which is limited to well-characterized species containing up to two carbon atoms (C2). \textit{FRECKLL} includes vertical mixing parameterized by a user-defined eddy diffusion coefficient ($K_{zz}$), and photolysis rates are computed using a two-stream approximation with stellar UV flux inputs. The model does not currently account for condensation processes, including \hho cloud formation, which may impact the abundance profiles of condensable species at lower temperatures. Since this could impact the retrieved C/O ratio, we performed a parallel fit to the observations excluding the \hho opacity, which biases the C/O ratio in the opposite direction.

\textit{FRECKLL} has been originally designed to model equilibrium and disequilibrium chemistry on warm to hot hydrogen-dominated exoplanetary atmosphere \citep{Venot2012,Venot2015,Venot2020a,Veillet2024}. However, it can also be applied to temperate atmospheres, with validation for temperatures ranging from 300 to 2500 K. We simulate the atmosphere of K2-18\,b for a range of C/O ratio (from 0.1 to 100), metallicity (from 0.1 to 1000) and constant eddy diffusion coefficient (K$_{zz}$ from 10$^5$ to 10$^{10}$ cm$^2$.s$^{-1}$), see Table \ref{tab: grid} for the grid range and resolution. To achieve a given C/O ratio, we adopt the approach of adjusting both carbon and oxygen abundances while conserving the overall metallicity.

\begin{table}[!htbp]
    \caption{Non-equilibrium chemical model running grid.}
    \centering
    \begin{threeparttable}
    \begin{tabular}{@{}lccc@{}}
    \toprule
    \toprule
    \textbf{Parameters}     & \textbf{range}  & \textbf{step} & \textbf{Npoint}   \\ \midrule
    log$_{10}$[Metallicity] & [-1:3]          & 0.137         & 30                \\
    C/O                     & [0.1:10]        & 0.341         & 30                \\
    C/O                     & [10:20]         & 0.5           & 20                \\
    C/O                     & [20:100]        & 10            & 8                 \\
    log$_{10}$[K$_{zz}$]    & [5:10]          & 1.0           & 6                 \\ \bottomrule
    \end{tabular}
    \tablefoot{The range, the number of points and the corresponding step are described. C/O have a higher resolution at lower values to better characterize the main variations and the lower limit. Total number of simulation are 30x58x6.}
    \end{threeparttable}
    \label{tab: grid}
\end{table}

We used the temperature profile from \cite{Blain2021}, corresponding to the best fit (metallicity = 175, C/O = 0.13, K$_{zz}$ = 10$^6$). We neglect, at first-order, temperature variations due to composition, as we do not expect significant changes in the overall trend. This is consistent with the findings of \cite{Blain2021}, where the impact was shown to be small. The parameters of K2-18\,b are listed in Table \ref{tab: k2par}. For photochemistry, we used MUSCLES GJ 436 version 2.2 stellar spectra \citep{France2016,Youngblood2016,Loyd2016} as a proxy for K2-18, as suggested by \cite{dosSantos2020}.\\

\begin{table}[!htbp]
    \caption{Planet-Stellar K2-18\,b parameters.}
    \centering
    \begin{threeparttable}
    \begin{tabular}{@{}lc@{}}
    \toprule
    \toprule
    \textbf{Parameters}     & \textbf{Values}     \\ \midrule
    R$_*$ [Rsun]            & 0.4445              \\
    T$_*$ [K]               & 3457                \\
    R$_p$ [Rjup]            & 0.2328              \\
    a [AU]                  & 0.1591              \\
    g [m/s$^2$]             & 12 (at 100bar)      \\ \midrule
    Reference               & \cite{Benneke2019}  \\ \bottomrule
    \end{tabular}
    \tablefoot{Used for \textit{FRECKLL} and \textit{TauREx} simulations.}
    \end{threeparttable}
    \label{tab: k2par}
\end{table}

\subsection{Fitting to observation}

With the atmospheric profiles output by \textit{FRECKLL}, we used the Forward Model framework of \textit{TauREx 3} \citep{Al-Refaie2021} to generate transmission spectra. We considered the same cross-sections and molecules as for the equilibrium chemistry retrieval.

We adjusted the theoretical spectra to match the K2-18\,b observational \textit{native resolution} data from \cite{Madhusudhan2023} by retrieving the planet's radius. Then, we computed the chi-square ($\chi^2$) to determine the best-fit models. As in \cite{Blain2021} and \cite{Wogan2024}, this approach provides an initial insight into the most accurate non-equilibrium chemical model compared to observations, though it has not yet been fully implemented within retrieval model frameworks. We estimate a good fit at 2$\sigma$.

We reduce the 3D parameter space (metallicity, C/O, K${zz}$) to 2D by selecting the minimum $\chi^2$ values along the K${zz}$ dimension, in order to explore the effects of metallicity and C/O. The best-fit model corresponds to the minimum $\chi^2$, with 1$\sigma$, 2$\sigma$, 3$\sigma$, and 4$\sigma$ uncertainties in a 2D parameter space, defined as a $\Delta\chi^2$ difference from the best fit, with values of 2.30, 6.18, 11.83, and 19.33, respectively. To refine our estimates, we fit (polynomial fit) the projected $\Delta\chi^2$ over the C/O ratio and metallicity, obtaining the best-fit values along with 1$\sigma$, 2$\sigma$ and 3$\sigma$ confidence intervals. The projected $\Delta\chi^2$ is obtained by reducing the 2D parameter space to 1D, selecting the models with the lowest $\chi^2$ values along the collapsed dimension. In a 1D parameter space 1$\sigma$, 2$\sigma$, 3$\sigma$, 4$\sigma$, and 5$\sigma$ uncertainties correspond to values of 1, 4, 9, 16, and 25, respectively.

This method provides an estimate of the C/O ratio and metallicity of K2-18\,b, which in turn constrains the abundances of minor species. Additionally, we fitted a flat-line model to the observed spectrum and we calculated the associated $\chi^2$ to assess the significance of our retrieved models, ensuring that they provide a better fit than scenarios such as a bare rock with no atmosphere or a gas giant with a cloud layer.


\section{Results}

Table \ref{tab: results} summarizes the key results for non-equilibrium chemical models. Metallicity, C/O ratio, and K$_{zz}$ were calculated using a polynomial fit to the global trend. Best-fit abundances and their $3\sigma$ uncertainties were derived from the explored model grid. The best-fit values (on the model grid) are 280.7 for metallicity, 90.0 for the C/O ratio, and $10^6$~cm$^2$\,s$^{-1}$ for K$_{zz}$, with a corresponding $\chi^2$ of 4786.25.

\begin{table}[!htbp]
    \caption{Retrieved parameters and species abundances of the best fit.}
    \centering
    \begin{threeparttable}
    \begin{tabular}{@{}lc@{}}
    \toprule
    \toprule
    \textbf{Parameters}       & \textbf{Values}                         \\ \midrule
    Metallicity (1$\sigma$)   &  266$^{+90}_{-61}$                      \\
    Metallicity (2$\sigma$)   &  266$^{+291}_{-104}$                    \\
    Metallicity (3$\sigma$)   &  $\ge$ 129                              \\
    C/O (1$\sigma$)           &  $\ge$ 7.1                              \\
    C/O (2$\sigma$)           &  $\ge$ 2.1                              \\
    C/O (3$\sigma$)           &  $\ge$ 0.96                             \\
    K$_{zz}$ [cm$^2$.s$^{-1}$]&  All within 1$\sigma$                   \\ \midrule
    \multicolumn{2}{c}{Abundances at 1 mbar of best fit at 3$\sigma$}   \\ \midrule
    log$_{10}$[CH$_4$]        &  -0.3$^{+0.1}_{-1.7}$                   \\ [0.2cm]
    log$_{10}$[CO$_2$]        &  -5.6$^{+4.9}_{-0.9}$                   \\ [0.2cm]
    log$_{10}$[CO]            &  -2.2$^{+1.8}_{-0.6}$                   \\ [0.2cm]
    log$_{10}$[H$_2$O]        &  -4.2$^{+3.5}_{-4.3}$                   \\ [0.2cm]
    log$_{10}$[NH$_3$]        &  -3.8$^{+0.6}_{-5.8}$                   \\ [0.2cm]
    log$_{10}$[H$_2$CO]       &  -12.6$^{+5.9}_{-1.1}$                  \\ [0.2cm]
    log$_{10}$[CH$_3$OH]      &  -9.6$^{+3.3}_{-4.0}$                   \\ [0.2cm]
    log$_{10}$[CH$_3$CHO]     &  -11.3$^{+4.0}_{-1.8}$                  \\ [0.2cm]
    log$_{10}$[C$_2$H$_2$]    &  -7.0$^{+4.0}_{-7.5}$                   \\ [0.2cm]
    log$_{10}$[C$_2$H$_4$]    &  -9.6$^{+8.6}_{-1.1}$                   \\ [0.2cm]
    log$_{10}$[C$_2$H$_6$]    &  -4.1$^{+2.2}_{-2.0}$                   \\ [0.2cm]
    log$_{10}$[N$_2$]         &  -1.4$^{+0.3}_{-0.5}$                   \\ \bottomrule
    \end{tabular}
    \tablefoot{It is calculated using \textit{FRECKLL} and \textit{TauREx} simulations compared to JWST data \citep{Madhusudhan2023}. The species abundances values are taken at 1 mbar, which is representative to the probed atmosphere.}
    \end{threeparttable}
    \label{tab: results}
\end{table}

\subsection{Transmission spectra}

\begin{figure*}[h!]
\centering
\includegraphics[width=1.0\textwidth]{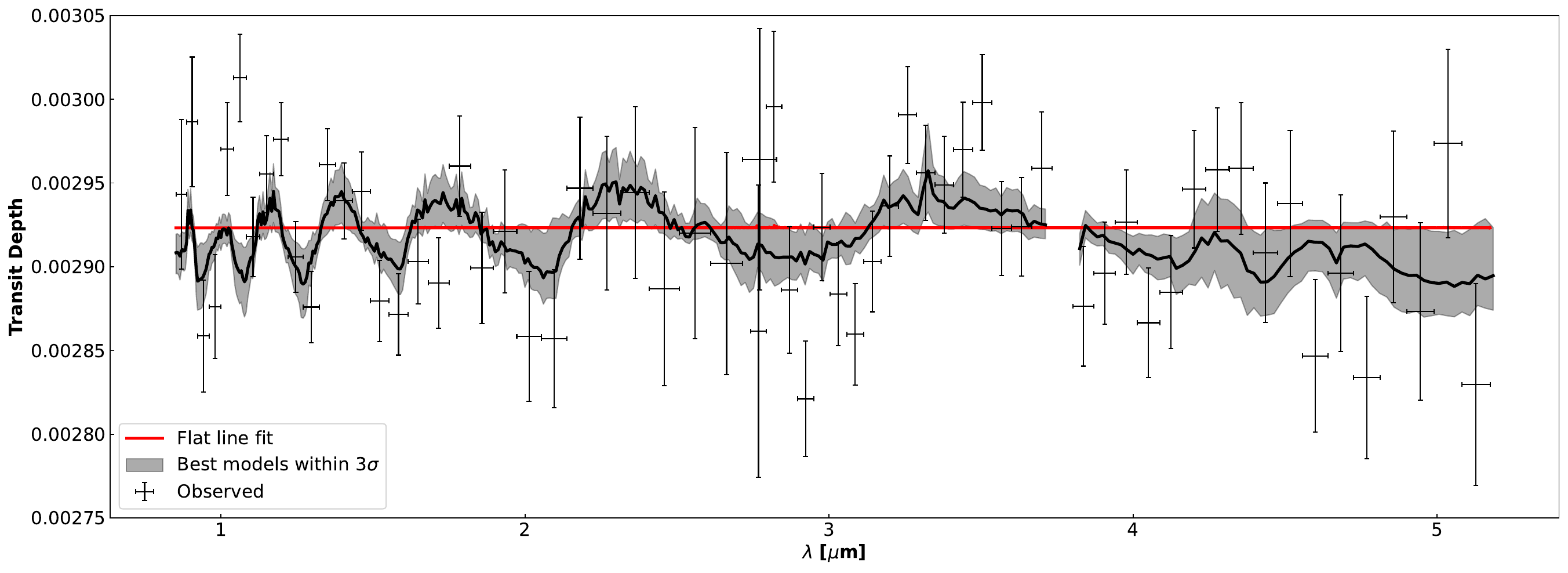}
\caption{Transit spectra of non-equilibrium 1D models at resolution 200, compare to K2-18\,b observation \citep{Madhusudhan2023} low resolution representation with an offset of -41ppm between NIRISS and NIRSpec data. Solid Gray line is the median value of the 3$\sigma$ dispersion from the best fit represented with the shaded area. Red line is the best flat line retrieval.}
\label{fig: spectra_all}
\end{figure*}

\begin{figure*}[h!]
\centering
\includegraphics[width=1.0\textwidth]{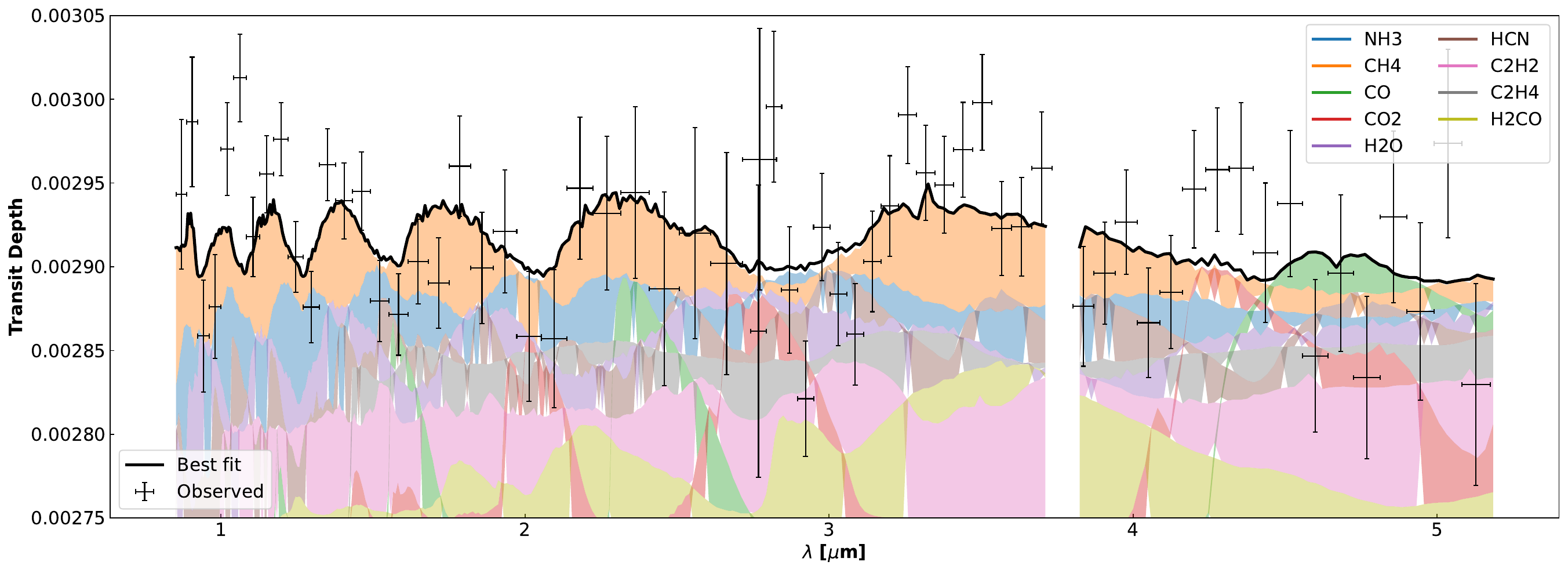}
\caption{Transit spectra of the best fit non-equilibrium 1D model at resolution 200 (black solid line). This is the approximate best fit on the simulated grid of metallicity, C/O and K$_{zz}$. Values are respectively 280.7, 90.0 and 10$^{6}$ cm$^2$.s$^{-1}$. It is compare to K2-18\,b observation \citep{Madhusudhan2023} low resolution representation with an offset of -41ppm between NIRISS and NIRSpec data. The contributions of considered molecules are represented with shaded colors.}
\label{fig: spectra_best}
\end{figure*}

The transmission spectra within 3$\sigma$ of the best fit are shown in Figure~\ref{fig: spectra_all}, with the shaded area representing the uncertainty. The main spectral features are attributed to CH$_4$, CO$_2$, and CO, while secondary features from H$_2$O and NH$_3$ can be observed near 2 and 3 $\mu$m (see Figure~\ref{fig: spectra_best}).

The significance of the disequilibrium models was assessed by comparing them to a flat-line retrieval (see Figure~\ref{fig: spectra_all}). The $\chi^2$ calculations (Figure~\ref{fig: chi_square_1d_zoom}) indicate that the best disequilibrium model achieves a confidence level exceeding 4$\sigma$ compared to the flat-line model. This confirms the robustness of the observed spectral features and the validity of the best disequilibrium model.

\subsection{Equilibrium chemistry retrieval}

The best fit is found for a metallicity of 71$^{+58}_{-20}$ and a C/O ratio of 6.78$^{+5.8}_{-2.0}$ (see Figure~\ref{fig: corner_plot} in Appendix \ref{an: equilibrium}). The spectral features are primarily attributed to CH$_4$ and CO$_2$ (see Figure~\ref{fig: spectra_best_eq} in Appendix \ref{an: equilibrium}). We observe little difference between equilibrium and non-equilibrium chemistry for the only strongly detected species, CH$_4$ (see Figures \ref{fig: spe_profile_eq} in Appendix \ref{an: equilibrium} and \ref{fig: spe_profile_best}). In fact, the calculated $\chi^2$ of the best-fit equilibrium chemical model is slightly lower than that of the best non-equilibrium model ($\Delta\chi^2 = -2.0$, so within 1$\sigma$ of the 2D space), indicating a marginally better fit with equilibrium chemistry, even though we know that the atmosphere is expected to be in a non-equilibrium chemical state.

Nevertheless, non-equilibrium chemistry significantly affects the abundances of H$_2$O, NH$_3$, and CO$_2$, which can either increase or decrease depending on metallicity and the C/O ratio. However, these variations remain below current detection limits.

\begin{figure}[h!]
\centering
\includegraphics[scale=0.4]{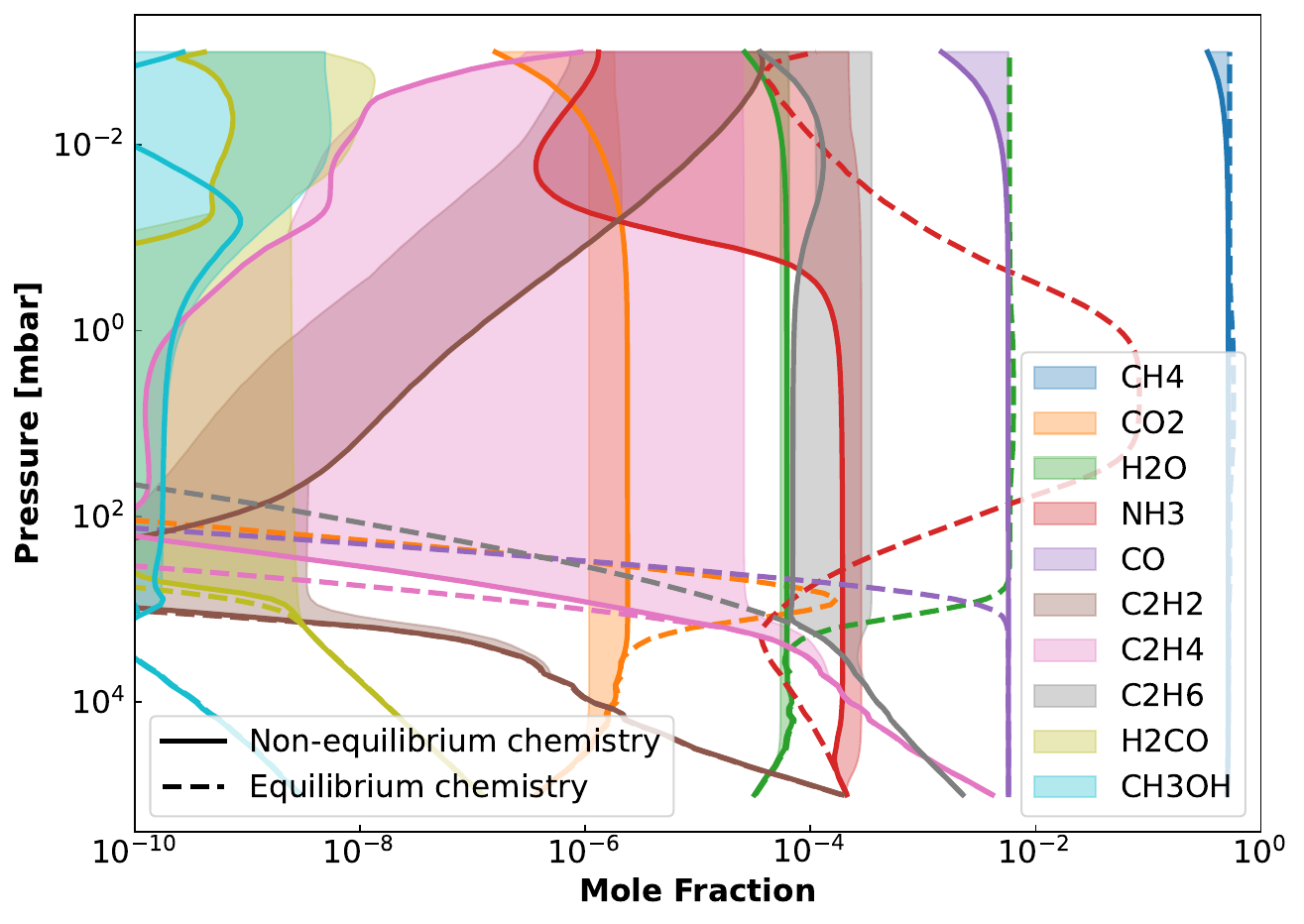}
\caption{Species profiles of the best fit of non-equilibrium 1D models (solid lines) and of equivalent equilibrium chemistry model (dash lines). This is the best fit on the models grid with metallicity of 280.7, C/O of 90.0, and K$_{zz}$ of 10$^6$ cm$^2$.s$^{-1}$. The shaded area represent the dispersion of K$_{zz}$ from 10$^6$ to 10$^{10}$ cm$^2$.s$^{-1}$.}
\label{fig: spe_profile_best}
\end{figure}

\subsection{Eddy diffusion coefficient}

The K$_{zz}$ is not constrained by our analysis. When reducing the 3D parameter space to 1D by selecting models with the lowest $\chi^2$ values across the metallicity and C/O dimensions, the resulting $\chi^2$ values remain within $1\sigma$ of the best fit. Near the best-fit metallicity, variations in K$_{zz}$ result in $\chi^2$ fluctuations of less than 1$\sigma$.

Figure~\ref{fig: spe_profile_best} shows the species profiles of the best-fit model. The shaded area represents the dispersion in K$_{zz}$. This illustrates that while vertical transport significantly affects minor species, it has little impact on major species, particularly \chhhh. This explains why K$_{zz}$ remains unconstrained in our analysis.

Figure~\ref{fig: chi_square_1d_zoom} presents the evolution of $\Delta\chi^2$ as a function of metallicity and C/O. Each parameter is projected along the other dimension using the lowest $\chi^2$ values. The shaded regions represent the dispersion due to variations in K$_{zz}$ over its entire range (from 10$^5$ to 10$^{10}$ cm$^2$.s$^{-1}$). The general trend is represented by the median value, which is used to determine the best-fit values or lower limits for metallicity and C/O through polynomial fitting, as shown in Figure~\ref{fig: chi_square_1d_zoom}. The corresponding numerical values are reported in Table~\ref{tab: results}.

\subsection{Metallicity}

\begin{figure}[h!]
\centering\includegraphics[scale=0.40]{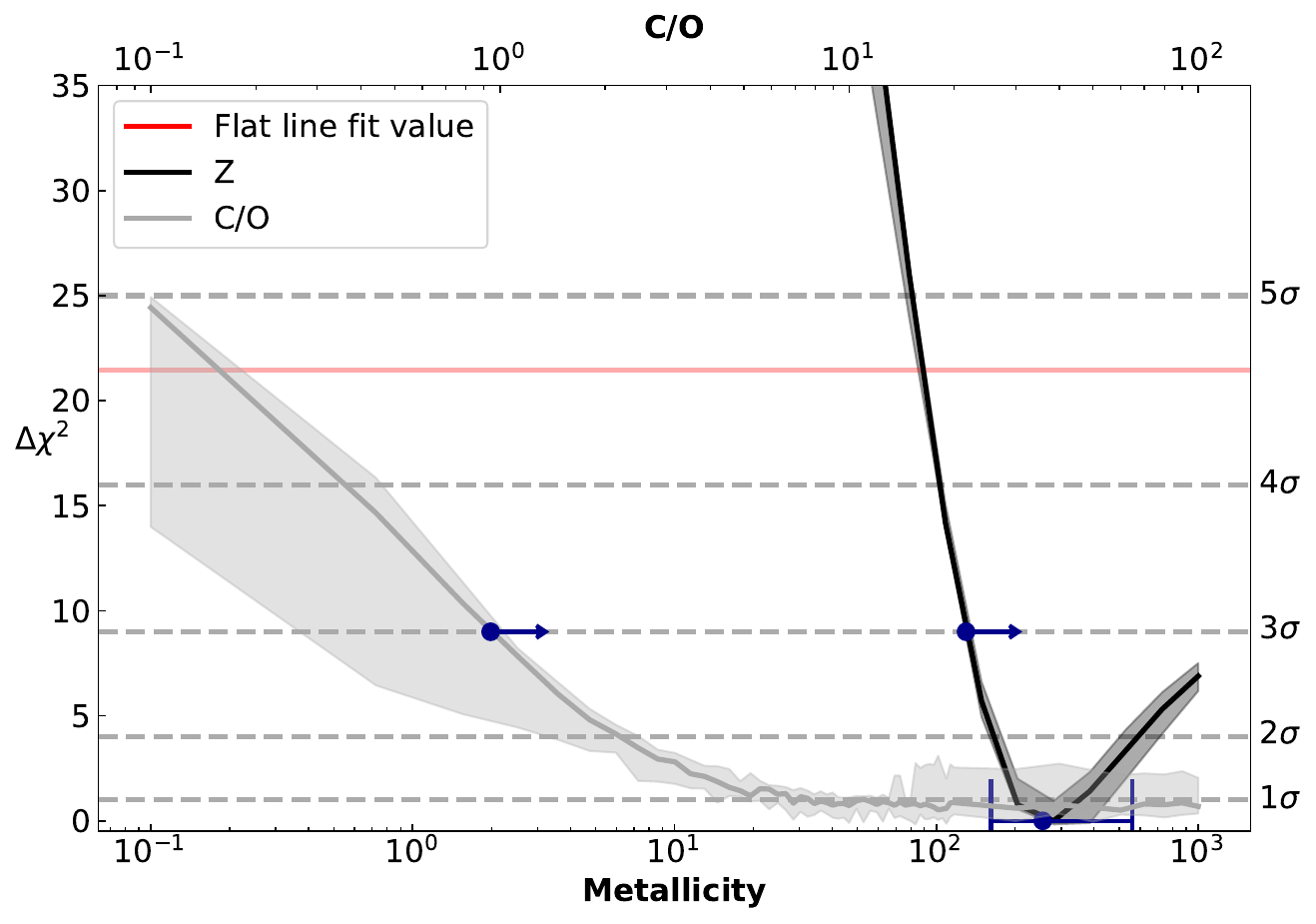}
\caption{Evolution of $\Delta\chi^2$ with metallicity or C/O ratio. The $\chi^2$ values are computed by comparing 1D chemical models to the K2-18\,b observations from \citet{Madhusudhan2023}. The $\Delta\chi^2$ values represent the difference from the best-fit model. The 1D projection is obtained by selecting the lowest $\chi^2$ across the other parameter dimension (metallicity or C/O). The shaded area represent the dispersion from K${zz}$ that have no preferred value (within 1$\sigma$), while the solid line indicates the median across K${zz}$ values. The red line represents the $\Delta\chi^2$ for the best flat-line retrieval, demonstrating a confidence level of the best fit exceeding 4$\sigma$. An additional best fit value with a 2$\sigma$ uncertainty for metallicity and a lower limit at 3$\sigma$ for metallicity and C/O are shown in blue, derived from a polynomial fit in log scale.}
\label{fig: chi_square_1d_zoom}
\end{figure}

\begin{figure}[h!]
\centering
\includegraphics[scale=0.43]{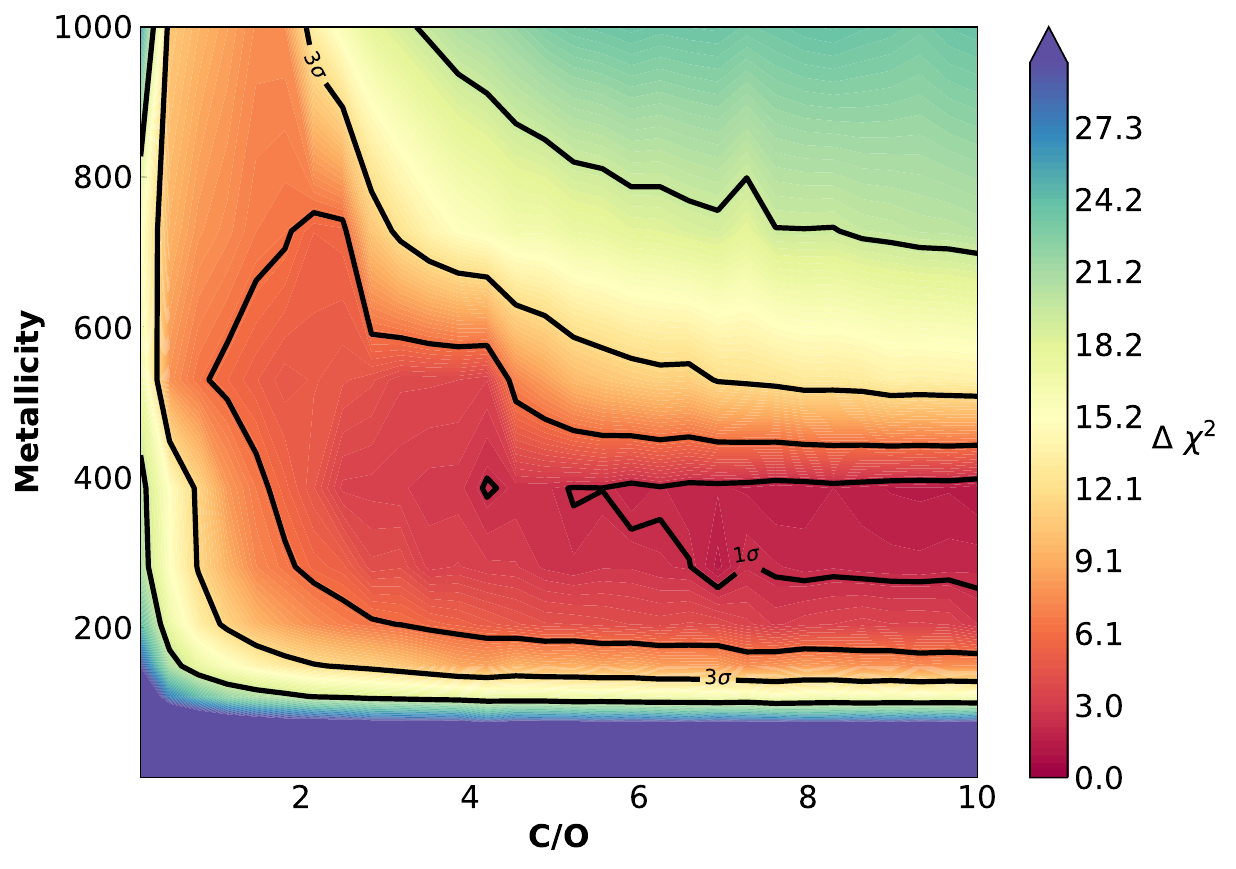}
\caption{Representation of the $\Delta\chi^2$ with the best fit compared to K2-18\,b observation \citep{Madhusudhan2023} and the 1D models. The 2D map goes over the range of metallicity and C/O ratio, collapsed in K$_{zz}$ with best $\chi^2$. The 2D map is zoomed for C/O ratio $\le$ 10 since above it follows the same trend.}
\label{fig: chi_square_2d}
\end{figure}

Metallicities $\lessapprox$ 100 are statistically less significant than a flat line, as shown in Figure~\ref{fig: chi_square_1d_zoom}. The best fit found in our model grid corresponds to a metallicity of 280.7. Using a polynomial fit, we derived a 2$\sigma$ confidence interval of 266$^{+291}_{-104}$, as shown in Figure~\ref{fig: chi_square_1d_zoom} and reported in Table \ref{tab: results}. Within our modeled grid, we could only establish a 3$\sigma$ lower limit of $\ge$129. Figure~\ref{fig: chi_square_2d}, which presents $\Delta\chi^2$ as a function of metallicity and C/O, shows that this is primarily due to the large uncertainty in metallicity at low C/O values. An upper limit at 3$\sigma$ might be obtained by extending the metallicity grid beyond 1000, though the benefit of doing so appears limited. Two distinct regimes can be identified:

\paragraph{C/O ratio $\lessapprox$ 2} High metallicity are favored ($\gtrapprox$ 100). But, the metallicity has huge uncertainties between 100 and 1000. 

\paragraph{C/O ratio $\gtrapprox$ 2} Increasing the C/O ratio constrains the metallicity between 205 and 356 at 1$\sigma$ of the best fit (see Figure~\ref{fig: chi_square_1d_zoom}). This follows \cite{Fortney2013} expectation considering the radius of the planet.\\

Figure~\ref{fig: chi_square_2d} is zoomed for C/O $\le$ 10 since until 100 it follows the same trend.

\subsection{C/O ratio}

Low C/O ratio, close to solar C/O ($0.55^{+0.130}_{-0.108}$ \cite{Lodders2019}), are statistically as significant as a flat line, as shown in Figure~\ref{fig: chi_square_1d_zoom}. Our results favor a high C/O ratio (see Figure~\ref{fig: chi_square_2d}). We are only able to constrain a lower limit on the C/O ratio: values above 7.1 lie within 1$\sigma$ of the best fit, while the solar C/O ratio is ruled out at 3$\sigma$ confidence level (see Figure~\ref{fig: chi_square_1d_zoom} and Table \ref{tab: results}). Excluding \hho opacity slightly changes the retrieved C/O ratio. The most affected values are at the 2$\sigma$ and 3$\sigma$ confidence levels, which decrease to 1.4 and 0.57, respectively, compared to the original values of 2.1 and 0.96. Accounting for \hho condensation would yield values that lie in between.

\subsection{Abundances}

\begin{figure*}[h!]
\centering
\includegraphics[scale=0.44]{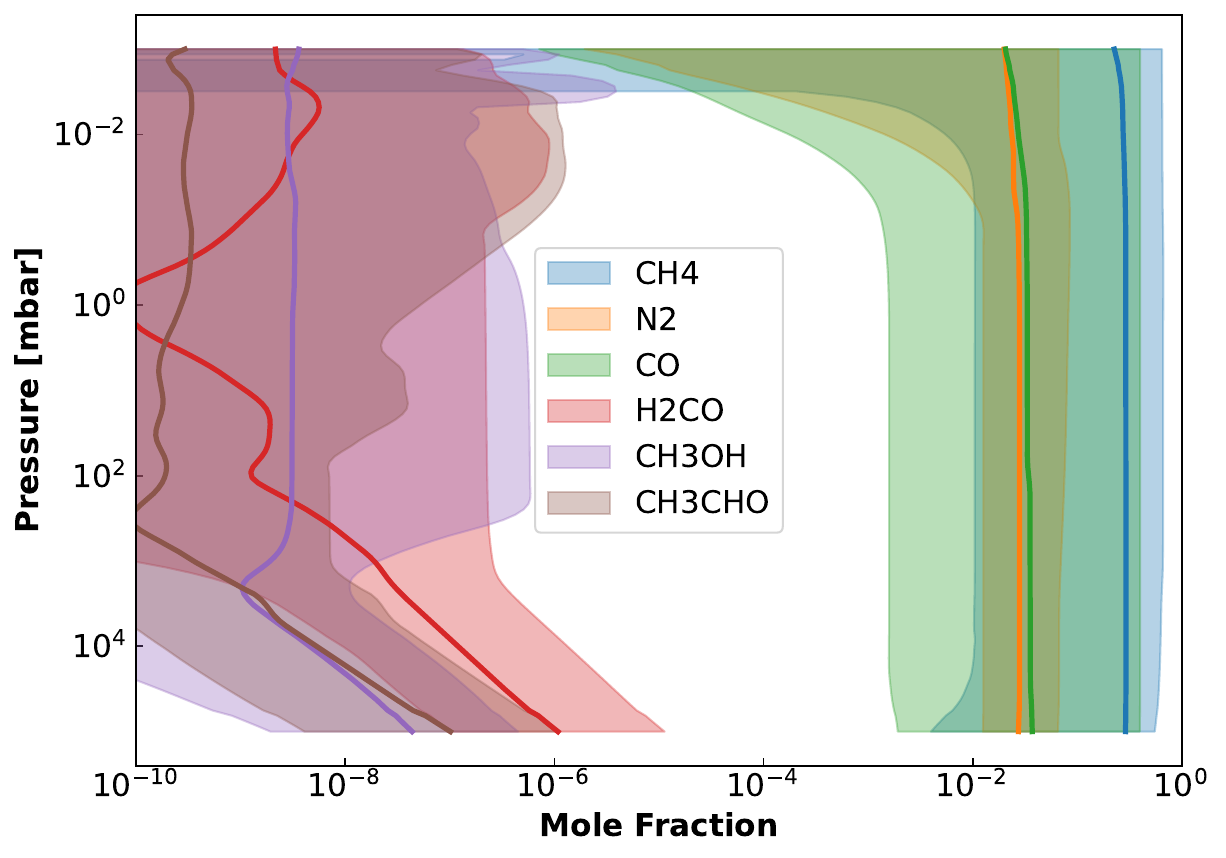}
\includegraphics[scale=0.44]{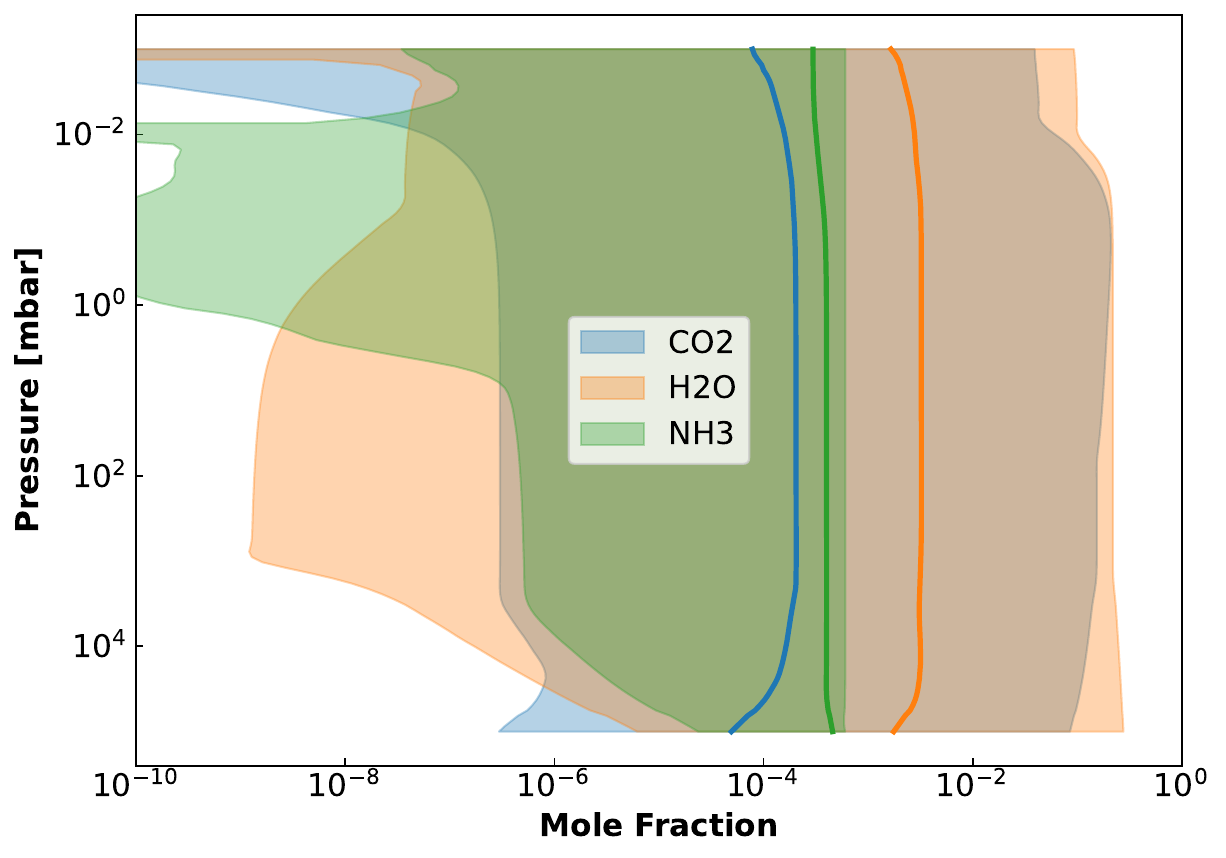}
\\
\includegraphics[scale=0.44]{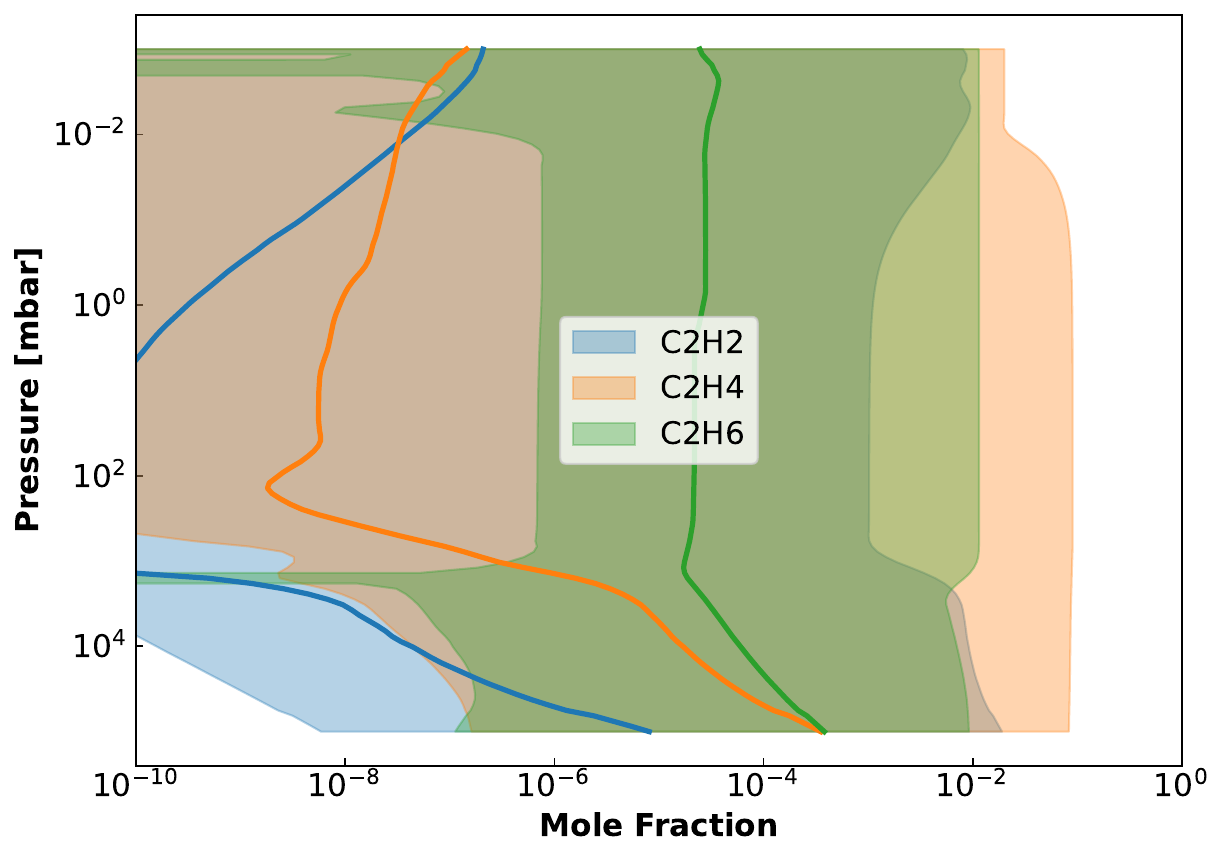}
\includegraphics[scale=0.44]{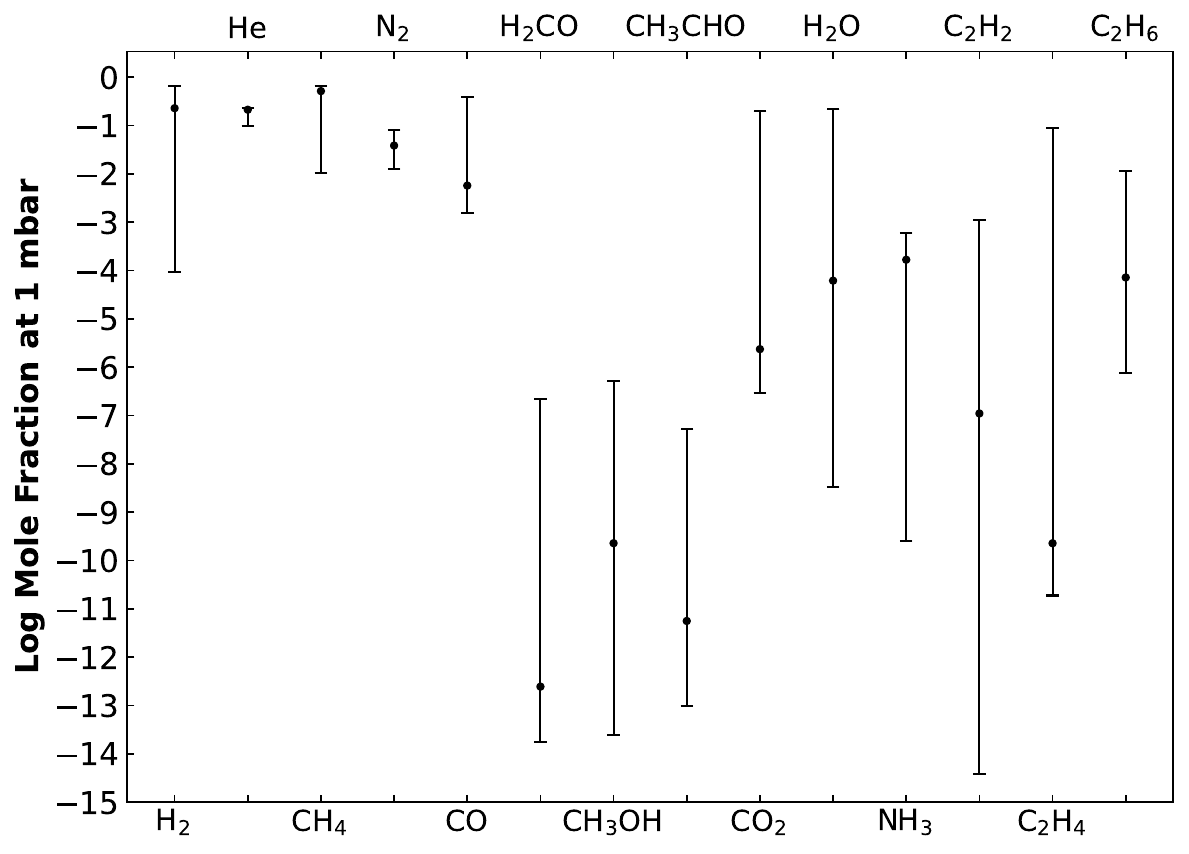}
\caption{Species profiles of non-equilibrium 1D models within 3$\sigma$ of the best fit, compare to K2-18\,b observation \citep{Madhusudhan2023}. Solid line is the median value of the dispersion represented with the shaded area. Bottom right shows values at 1 mbar with a dot for the best fit model on the parameter space grid.}
\label{fig: spe_profil}
\end{figure*}

Figure~\ref{fig: spe_profil} shows the profiles of key species within the 3$\sigma$ uncertainty range. The abundances at 1 mbar for the best-fit model, along with their 3$\sigma$ uncertainties, are also shown Figure~\ref{fig: spe_profil} and reported in Table \ref{tab: results}. This pressure level corresponds to the region probed by the observations, where the abundance profiles are shown also to be relatively vertically constant in the model.

At 3$\sigma$, CH$_4$, CO$_2$, CO, H$_2$O, and C$_2$H$_4$ can reach mixing ratios of up to 10\% (see Figure~\ref{fig: spe_profil}). However, CO$_2$, H$_2$O, and C$_2$H$_4$ can also fall below the ppm level, meaning we cannot confidently claim a detection for these species. In contrast, CH$_4$ is well constrained, with an abundance consistently above 1\%.

Additionally, oxidized species such as H$_2$CO, CH$_3$OH, and CH$_3$CHO can reach mixing ratios close to the ppm level (see Figure~\ref{fig: spe_profil}). While CH$_4$ remains the only strongly confirmed detection, a variety of chemically interesting species may be present in significant amounts.

\section{Discussion}

\subsection{Species detection}

To accurately characterize the atmosphere of K2-18\,b, particularly under its specific temperature, irradiation, and vertical transport conditions, it is essential to use comprehensive chemical models that incorporate non-equilibrium processes. The motivation for using such models arises not only from their better agreement with retrieved atmospheric abundances but also from the limitations imposed by the low signal-to-noise ratio (SNR) of current observations, which limits our ability to constrain the full atmospheric composition. While the atmosphere of K2-18\,b is likely chemically rich, retrievals assuming constant abundance profiles can confidently confirm only the presence of \chhhh.

The \chhhh\ and \coo\ detections with inferred abundances of $\sim$1\% reported by \cite{Madhusudhan2023} are within $3\sigma$ of our best-fit model. However, as shown in Figure~\ref{fig: spe_profil}, \chhhh\ appears more abundant than \coo, an observation consistent with disequilibrium models such as those by \cite{Wogan2024}. Moreover, our analysis shows that \coo\ abundances can drop as low as log$_{10}$[\coo] = -6.5 at 1\,mbar (Table~\ref{tab: results}), aligning with the findings of the reanalysis by \cite{Schmidt2025}. The large uncertainty in \coo\ abundance retrieved with non-equilibrium models further supports the conclusion of \cite{Schmidt2025}, who, unlike \cite{Madhusudhan2023}, question the robustness of the \coo\ detection.

Indeed, the updated data reduction from \cite{Schmidt2025} limits confirmed detections to \chhhh, casting further doubt on the presence of \coo. This is evident from the increased uncertainty around the 4.2\,$\mu$m \coo\ band and the wide range of retrieved abundances in our analysis (see Figure~\ref{fig: spe_profil}). Nevertheless, this does not rule out the presence of \coo\ or other minor species. As shown in Figure~\ref{fig: spectra_best}, several of these compounds may still contribute to the observed spectrum and could exist in significant quantities, albeit below current detection thresholds.

Other species such as CO, \hho, and \cchhhh\ can reach mixing ratios of a few percent but remain undetected, likely because their spectral signatures are either masked by \chhhh\ or buried in observational noise. Our results, for instance, indicate substantial CO abundances (Figure~\ref{fig: spe_profil} and Table~\ref{tab: results}) with a notable spectral feature near 5\,$\mu$m (Figure~\ref{fig: spectra_best}), despite CO not being detected in previous studies. This non-detection likely stems from large data uncertainties affecting CO’s spectral features, an issue that similarly impacts other secondary species like \hho\ and \nhhh.

\nhhh, in particular, can reach log$_{10}$[\nhhh] = -3.2 at 1\,mbar (Table~\ref{tab: results}), consistent with predictions from disequilibrium models such as \cite{Wogan2024}, and contributes to secondary features in the spectrum (Figure~\ref{fig: spectra_best}). Unlike \cite{Madhusudhan2023} and \cite{Schmidt2025}, who reported upper limits of log$_{10}$[\nhhh] = -4.51 and -5.25 respectively, our disequilibrium model accommodates higher \nhhh\ abundances without requiring nitrogen depletion, challenging the interpretation by \cite{Schmidt2025}. The apparent absence of \nhhh\ in the spectra is more consistent with a gas mini-Neptune scenario rather than the purely Hycean interpretation proposed by \cite{Madhusudhan2023}. It is important to note, however, that nitrogen may be sequestered through dissolution in deep magma oceans, as proposed by \cite{Shorttle2024}, thereby lowering its atmospheric abundance. Consequently, the upper bound of log$_{10}$[\nhhh] $\lesssim$ –3.2 should not be interpreted as a strict observational constraint but rather as a model-limited outcome under the assumption of undepleted nitrogen. Additionally, since our approach assumes a fixed P–T profile, the retrieved abundance of \nhhh\ is subject to uncertainty, and should be interpreted as indicative of chemical trends rather than precise values.

Similarly, the atmospheric H$_2$O abundance predicted by our models may be biased high, since \textit{FRECKLL} does not include condensation processes. In K2-18\,b’s cool upper atmosphere, H$_2$O vapor could condense and rain out, effectively reducing the gas-phase abundance in the observable layers. Despite these biases favoring higher-than-realistic abundances of \nhhh\ and \hho, neither species is robustly detected in the current data. Their spectral features appear to be masked by dominant absorbers such as \chhhh\ or buried in observational noise. This suggests that even if these molecules were present at relatively high levels, current observations lack the sensitivity and spectral resolution to confirm their presence.

As also seen in Figure~\ref{fig: spe_profil}, although the amplitude of spectral features in the best-fit models may appear smaller than in the observed data, these differences remain within the observational uncertainties. A key source of this mismatch is the lack of accurate model fitting at shorter wavelengths ($\leq 2\,\mu$m), where a decreasing trend, likely caused by aerosols, is not captured. This limitation leads to challenges in simultaneously fitting the 1–2\,$\mu$m and 2–5\,$\mu$m spectral ranges. As a result, models with pronounced features may overfit one region while underfitting the other. This discrepancy could also affect the retrieved C/O ratio, as the CO and \coo\ bands can be sensitive to the global shift and may be misrepresented without proper aerosol modeling.

A major challenge in atmospheric retrieval is that dominant spectral features, especially those of \chhhh, can obscure the signatures of other potentially abundant species such as \hho\ and \nhhh\ (see Figure~\ref{fig: spectra_best}). To resolve these overlapping signals, high-resolution ground-based spectroscopy offers a promising approach. Such observations could enable the detection of minor species, such as \hhco, \chhhoh, and \chhhcho, which may exist at ppm levels (see Figure~\ref{fig: spe_profil} and Table~\ref{tab: results}). While K2-18\,b remains a challenging target, upcoming instruments on next-generation telescopes, such as ANDES on the ELT, are expected to achieve the SNR necessary to detect hidden species like \hho\ and \nhhh \citep{Palle2025}, helping refine our understanding of K2-18\,b’s chemical inventory.

\subsection{Bimodality of the metallicity}

In contrast to the uncertainties surrounding other parameters, all analyses consistently support a high atmospheric metallicity for K2-18\,b, exceeding 100 times the solar value. However, when the C/O ratio is assumed to be near the solar value, metallicities up to nearly 1000 still fall within the $1\sigma$ confidence interval. This trend was already observed in HST data by \cite{Bezard2022}, which underlines the fact that the entire spectrum is dominated by \chhhh\ absorption bands, consistent with \chhhh\ being the only species robustly detected. This uncertainty aligns with the observation that, at low C/O ratios, the $\Delta\chi^2$ values become comparable to that of a flat line (see Figure~\ref{fig: chi_square_1d_zoom}), indicating that high-metallicity models, which naturally produce flatter spectra, can still yield statistically acceptable fits.

The implications of such high metallicity have been tempered in previous studies by invoking the formation of clouds, which become more likely at elevated metallicities due to the increased partial pressure of \hho. These clouds can further flatten the spectrum, as discussed in \cite{Charnay2021} and \cite{Blain2021}. However, these interpretations did not account for the potential effects of elevated C/O ratios.

In this study, by modeling an extensive grid spanning both metallicity and C/O ratio, we place tighter constraints on the atmospheric metallicity. Our results indicate that the metallicity is likely between 162 and 557 at the $2\sigma$ level (Table~\ref{tab: results}). Metallicities above $\sim$500 become inconsistent with the data when paired with high C/O ratios (in contrast to low C/O ratios $\lesssim 2$). This is because, while high metallicity flattens the spectrum, it also leads to \cchhhh\ dominating over \chhhh, resulting in a poorer match to the observed features.

The observed bimodality in the metallicity distribution, between low and high C/O regimes, emerges only because we explore the full joint parameter space of metallicity and C/O ratio. Studies assuming a fixed solar C/O ratio, as done in previous work \citep{Charnay2021,Bezard2022,Wogan2024}, exclude only low metallicities (i.e., $\lesssim 100$). By extending the exploration to elevated C/O ratios for the first time, we are able to observe this bimodality and place significantly stronger constraints on the metallicity of K2-18\,b.

\subsection{A high C/O ratio}

\cite{Madhusudhan2023} predict a low C/O ratio, based on the detection of both \coo and \chhhh at $\sim$1\%. This interpretation is used to favor a Hycean world over a deep atmosphere scenario, given the absence of detectable \nhhh and CO features, which should be present in significant quantities in a deep atmosphere with a low C/O ratio. However, the free retrieval approach employed in their study relies primarily on the dominant spectral features to estimate the C/O ratio. In contrast, our results show that \nhhh and CO can exist in substantial amounts without being detected, due to the high noise and large uncertainties in the data, especially when exploring scenarios with higher C/O ratios (see Figure~\ref{fig: spectra_best}). Such high C/O ratio scenarios have not been explored in previous studies, whereas our analysis favors a high C/O ratio (see Figure~\ref{fig: chi_square_2d}), consistent with the conclusions of \cite{Schmidt2025}.

While \cite{Schmidt2025} suggested a high C/O ($\gtrsim$3) based on the 2$\sigma$ uncertainty in the retrieved H$2$O abundance (log$_{10}$[H$_2$O] < –3.05), our analysis is based on the full spectral contributions. Nonetheless, our results are consistent within the 1$\sigma$ to 2$\sigma$ lower limits reported in Table \ref{tab: results}. However, as our approach assumes a fixed P–T profile, the derived constraints on the C/O ratio carry additional uncertainties, and should be viewed as indicative of chemical trends rather than precise determinations. Additionally, we demonstrated that the absence of \hho condensation in the model introduces a bias. The retrieved C/O ratio should give slightly reduce value, but not by more than 1.4 (at 2$\sigma$).

This unusually explored and finding of a high C/O ratio in an \hh-dominated atmosphere is therefore consistent with the observed \chhhh-rich spectrum. A carbon-rich atmosphere would naturally enhance aerosol formation, which may contribute to the spectral slope observed at wavelengths $\leq 2,\mu$m \citep{Madhusudhan2023}, a feature not captured by our models. However, current constraints on aerosols remain limited \citep{Madhusudhan2023,Schmidt2025}.

The C/O ratio cannot be reliably constrained using free retrieval methods alone and instead requires comprehensive disequilibrium chemical models for accurate inference. Even with such models, our results show that the C/O ratio is only constrained with a lower limit. This limitation stems from the current observational data, which do not provide strong constraints on species particularly sensitive to the C/O ratio. Consequently, this calls into question the reliability of using the C/O ratio alone to infer atmospheric chemistry. As noted by \citet{Turrini2021}, breaking the degeneracy in the C/O ratio, especially for giant planets, may require additional elemental ratios such as C/N, N/O or even S/N. \cite{Kama2019} and \cite{Kama2025} also suggested S/H, S/O and P/H ratio. However, in our case, there are no strong constraints on nitrogen-bearing species, limiting our ability to derive meaningful estimates of the C/N or N/O ratios, and the chemical network does not contain any S and P species.

At present, no single elemental ratio reliably constrains disequilibrium models, and, by extension, the overall atmospheric composition, without better constraints on additional species beyond \chhhh. As shown in Figure~\ref{fig: spectra_best}, both \coo and CO exhibit strong spectral features between 4 and 5,$\mu$m, but these features are poorly constrained due to the low SNR in this wavelength range. Since the abundances of CO and \coo are critical for constraining the C/O ratio, further transit observations with NIRSpec G395H would be particularly valuable for resolving this issue and advancing our understanding of the atmospheric composition of K2-18\,b.

\subsection{Unconstrained vertical mixing in the atmosphere}

The variation of the eddy diffusion coefficient (K$_{zz}$) has no significant impact on the model fits and remains within the 1$\sigma$ uncertainty range across the best-fit solutions within the 3$\sigma$ confidence interval for metallicity. Similar conclusions were drawn by \cite{Blain2021} and \cite{Bezard2022}. This insensitivity is primarily due to the lack of observational constraints on species that are strongly influenced by variation of vertical quenching, which is governed by the transport efficiency parameterized by K${zz}$. In particular, the \chhhh abundance profile remains nearly constant and shows little dependence on K$_{zz}$ within the explored metallicity range and pressure levels (see Figures \ref{fig: spe_profil} and \ref{fig: spe_profile_best}). Thus, although vertical transport causes notable deviations from chemical equilibrium, variations in K$_{zz}$ themselves do not significantly constrain the data; rather, they introduce additional uncertainty in the retrieved abundances of minor species.

\subsection{What can we really learn from the observational data?}

Finally, many strong conclusions remain limited due to the large uncertainties in the observational data. Even when using self-consistent models with non-equilibrium chemistry, we are constrained by the lack of sufficient data precision. Given that only one molecule, \chhhh, is robustly detected, no meaningful upper limit on the C/O ratio can be determined. We further demonstrate, using a Bayesian retrieval with an equilibrium model, that the presence of non-equilibrium chemistry in the atmosphere of K2-18\,b cannot be conclusively proven. In fact, the best-fit equilibrium model is statistically more significant than the non-equilibrium model, with a $\Delta\chi^2 = -2$.

Nevertheless, we can still confidently conclude the detection of \chhhh, a correspondingly high C/O ratio of $\geq 2.1$ at the 2$\sigma$ level, and a metallicity of 266$^{+291}_{-104}$ at 2$\sigma$, acknowledging that these retrieved values are subject to model uncertainties due to the use of a fixed P–T profile.


\section{Conclusion}

In this study, we applied non-equilibrium chemical models to gain deeper insights into the atmosphere of K2-18\,b, a temperate sub-Neptune located within the habitable zone of an M-dwarf star. By combining the latest observational data from the JWST with the \textit{FRECKLL} chemical model, we explored the planet’s atmospheric composition and the underlying chemical processes in detail. Our analysis provides a comprehensive view of the key parameters (metallicity, C/O ratio, and K$_{zz}$) that shape the atmospheric composition of K2-18\,b. We explored for the first time such high C/O ratio (from 0.1 to 100) showing the importance exploring simultaneously those 3 key parameters. We warn that accurately constraining the planet's atmospheric composition requires more than traditional retrieval models assuming constant abundances; it also necessitates non-equilibrium models to address the limited species constraints imposed by the low SNR of the data.

Our results suggest a high atmospheric metallicity, exceeding 100 times the solar value. This finding aligns with previous studies, supports the planet’s potential for complex atmospheric chemistry, and constrains the metallicity to be below 558 at 2$\sigma$, acknowledging model uncertainties due to the use of a fixed P–T profile. We confirm that the presence of \chhhh is robustly detected, while the detection of \coo remains uncertain due to significant observational noise and large uncertainties. The abundances of minor species such as CO, \hho, and \nhhh, although not directly detectable, are predicted to be present in notable quantities, suggesting the chemical diversity of the atmosphere. These findings underscore the importance of non-equilibrium models for accurately interpreting exoplanetary atmospheres, especially in the case of temperate sub-Neptunes, where disequilibrium processes likely play a significant role. Non-equilibrium models offer a powerful tool to go beyond the limits of detectability, enabling us to investigate the hidden chemical processes and probe the plausible abundances of undetected species.

Our study also highlighted the challenges in constraining the C/O ratio with the available data. While we established a lower limit for the C/O ratio of 2.1 at the 2$\sigma$ level, we cannot derive any specific value overall, pointing to the need for higher-quality data. However, the high C/O ratio and the apparent decreasing trend at lower wavelength $\le$ 2$\mu$m can suggest the presence of hazes. The variation of eddy diffusion (K$_{zz}$) was found to have minimal impact on our results, consistent with previous studies, due to the lack of strong constraints from species influenced by variation of vertical mixing.

In conclusion, K2-18\,b’s atmosphere is likely to be chemically rich and complex. Our work lays the foundation for future studies aimed at better understanding the atmospheric dynamics, chemistry, and potential habitability of temperate sub-Neptunes, which remain a promising class of exoplanets for future investigations into the possibility of life beyond our solar system.

In perspective, future observations with JWST's NIRSpec G395H mode will significantly improve the SNR in the 4–5$\mu$m range, enabling more precise constraints on CO and CO$_2$, which exhibit strong spectral features in this wavelength region and are key tracers of the atmospheric C/O ratio. Complementarily, high-resolution spectroscopy with ELT/ANDES will provide unprecedented sensitivity and spectral resolution, facilitating the detection of spectrally blended or weak features from major species such as H$_2$O and NH$_3$. These molecules are crucial not only for characterizing atmospheric composition but also for constraining elemental ratios such as C/O and N/H. Together, these next-generation facilities will give more robust constraints on atmospheric non-equilibrium chemistry. This synergy between space- and ground-based platforms will be essential for a more comprehensive characterization of temperate sub-Neptune atmospheres like that of K2-18\,b.


\begin{acknowledgements}
This project has received funding from the European Research Council (ERC) under the ERC OxyPlanets projects (grant agreement No. 101053033). This project acknowledges funding from the European Research Council (ERC) under the European Union's Horizon 2020 research and innovation programme (grant agreement No. 679030/WHIPLASH). O.V. acknowledges funding from the ANR through the project ‘EXACT’ (ANR-21-CE49-0008-01), as well as from the Centre National d’\'Etudes Spatiales (CNES). We would like to thank the anonymous referee for his pertinent comments, which improved the presentation of our results.
\end{acknowledgements}

\bibliographystyle{aa}
\bibliography{biblio}

\begin{appendix}

\onecolumn

\section{Equilibrium chemistry retrieval results}
\label{an: equilibrium}

In this appendix, we present all the results from the complementary retrieval assuming equilibrium chemistry. Although K2-18\,b is expected to exhibit non-equilibrium chemistry, these results show that a single transit observed with NIRISS and NIRSpec G395H is insufficient to distinguish between equilibrium and non-equilibrium chemical states. This limitation arises from the lack of constraints on most species, with the exception of \chhhh, which is not strongly sensitive to non-equilibrium processes in K2-18\,b’s atmosphere.

\begin{figure*}[h!]
\centering
\includegraphics[width=1.0\textwidth]{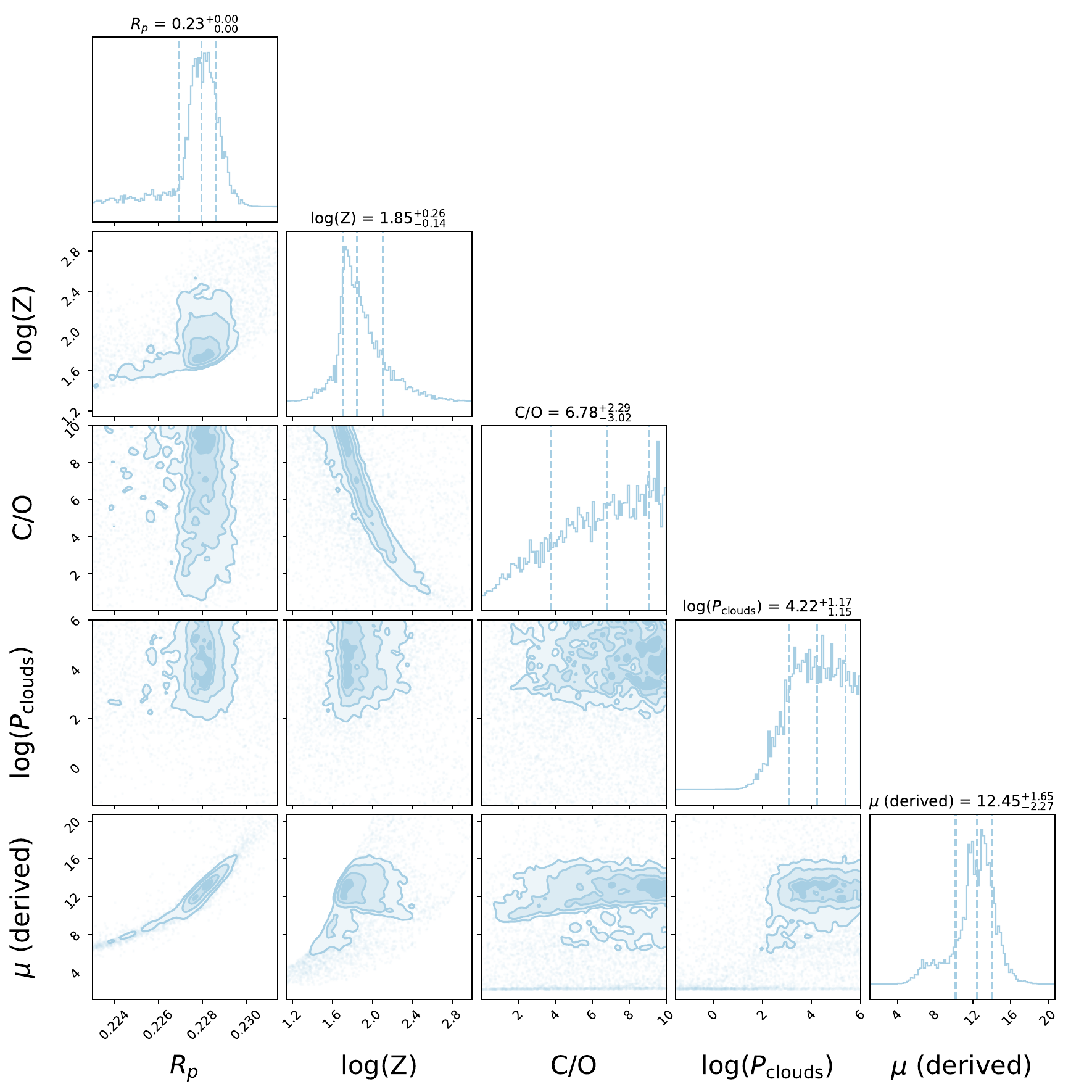}
\caption{Posterior of the retrieval performed with equilibrium chemistry on the K2-18\,b observation \citep{Madhusudhan2023} native resolution with an offset of -41ppm between NIRISS and NIRSpec data.}
\label{fig: corner_plot}
\end{figure*}

\begin{figure*}[h!]
\centering
\includegraphics[width=1.0\textwidth]{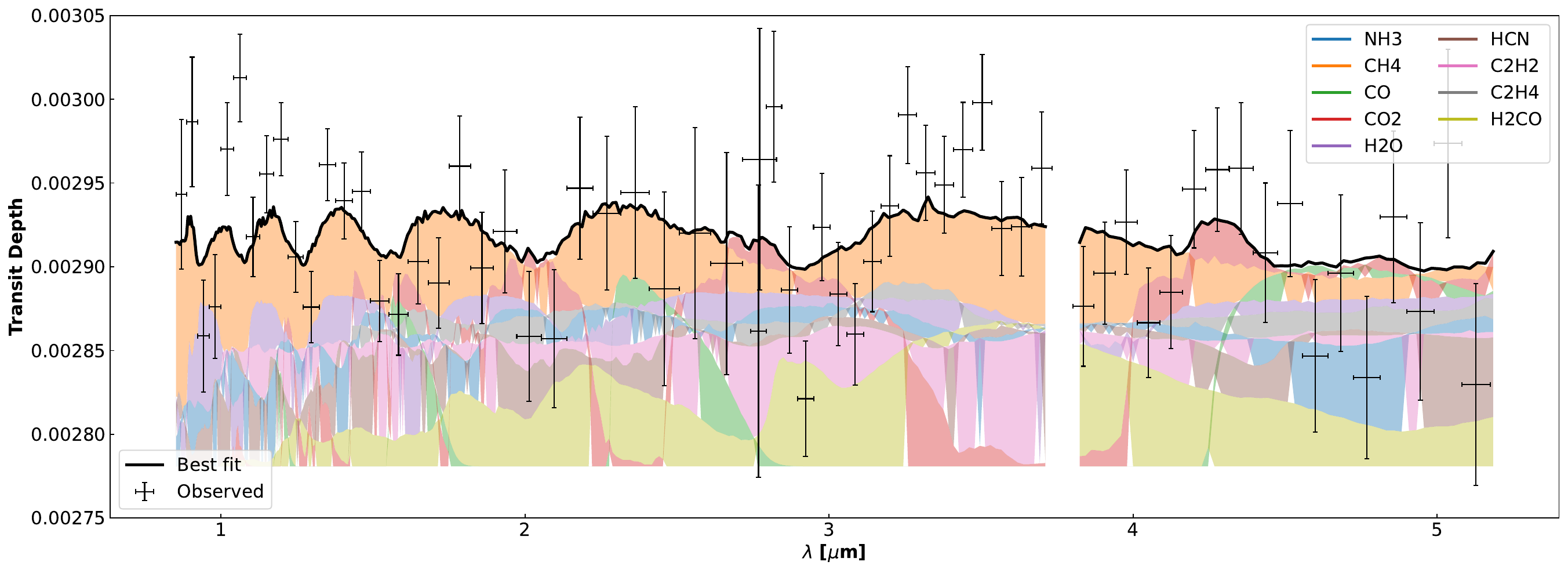}
\caption{Transit spectra of the retrieved best fit equilibrium at resolution 200 (black solid line). It is compare to K2-18\,b observation \citep{Madhusudhan2023} low resolution representation with an offset of -41ppm between NIRISS and NIRSpec data. The contributions of considered molecules are represented with shaded colors.}
\label{fig: spectra_best_eq}
\end{figure*}

\begin{figure*}[h!]
\centering
\includegraphics[width=1.0\textwidth]{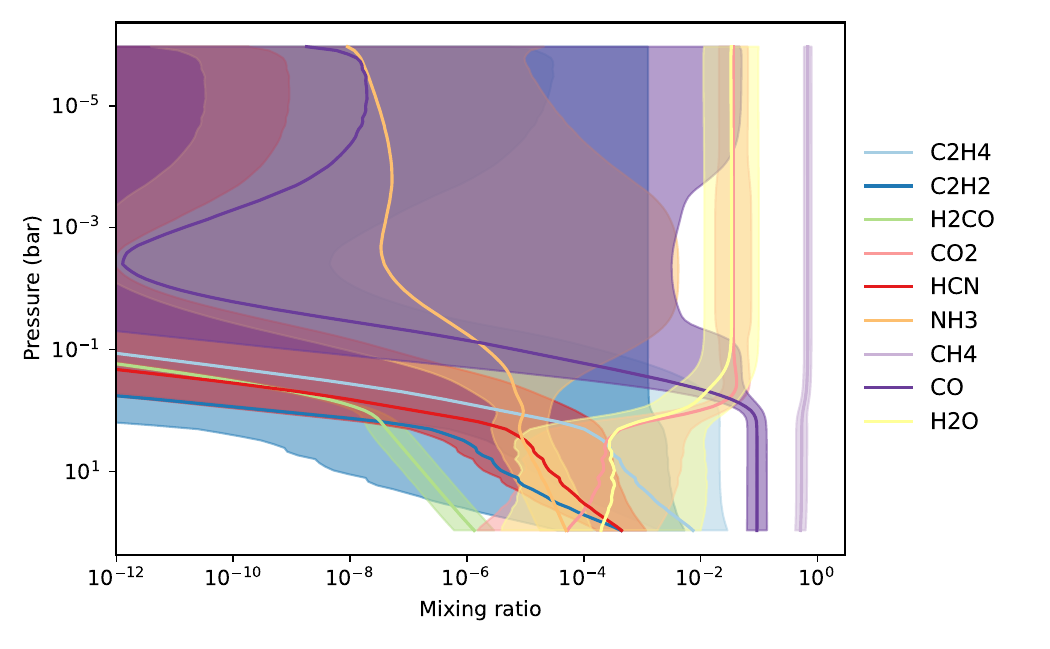}
\caption{Species profiles of the retrieved best fit (solid line) with its uncertainty (shaded area).}
\label{fig: spe_profile_eq}
\end{figure*}

\end{appendix}
%
%

\end{document}